\documentclass[%
aps,prc,twocolumn,nofootinbib,floatfix,superscriptaddress ]{revtex4-2}

\usepackage{graphicx}
\usepackage{dcolumn}
\usepackage{bm}

\usepackage{multirow}
\usepackage{amssymb}
\usepackage{amsmath, bbm}
\usepackage[figuresright]{rotating}
\usepackage{capt-of}

\newcommand{\La}{{\Lambda}}
\newcommand{\Si}{{\Sigma}}

\newcommand{\be}{\begin{eqnarray}}
\newcommand{\ee}{\end{eqnarray}}

\newlength{\feynwidth} 
\newlength{\feynwidthbig} 

\usepackage{xcolor} 

\usepackage{hyperref}
\usepackage{graphicx}

\hypersetup{colorlinks=true,citecolor= blue}

\usepackage[english]{babel}
\usepackage[autolanguage]{numprint}

\begin{document}


\title{Ab initio calculation of charge symmetry breaking in 
\texorpdfstring{$A=7$}{A=7}  and 8 \texorpdfstring{$\Lambda$}{Lambda}-hypernuclei }

\author{Hoai Le}%
\email{h.le@fz-juelich.de}
\affiliation{Institut f\"ur Kernphysik, Institute for Advanced Simulation 
and J\"ulich Center for Hadron Physics, Forschungszentrum J\"ulich, 
D-52425 J\"ulich, Germany}

\author{Johann Haidenbauer }%
\email{j.haidenbauer@fz-juelich.de}
\affiliation{Institut f\"ur Kernphysik, Institute for Advanced Simulation 
and J\"ulich Center for Hadron Physics, Forschungszentrum J\"ulich, 
D-52425 J\"ulich, Germany}

\author{Ulf-G. Mei{\ss}ner }%
\email{meissner@hiskp.uni-bonn.de}
\affiliation{Helmholtz-Institut~f\"{u}r~Strahlen-~und~Kernphysik~and~Bethe~Center~for~Theoretical~Physics,
~Universit\"{a}t~Bonn,~D-53115~Bonn,~Germany}
\affiliation{Institut f\"ur Kernphysik, Institute for Advanced Simulation 
and J\"ulich Center for Hadron Physics, Forschungszentrum J\"ulich, 
D-52425 J\"ulich, Germany}
\affiliation{Tbilisi State University, 0186 Tbilisi, Georgia}
\affiliation{CASA, Forschungszentrum J\"ulich, D-52425 J\"ulich, Germany}

\author{Andreas Nogga}
\email{a.nogga@fz-juelich.de}
\affiliation{Institut f\"ur Kernphysik, Institute for Advanced Simulation 
and J\"ulich Center for Hadron Physics, Forschungszentrum J\"ulich, 
D-52425 J\"ulich, Germany}
\affiliation{CASA, Forschungszentrum J\"ulich, D-52425 J\"ulich, Germany}

\date{October 2022}

\begin{abstract}
The $\Lambda$ separation energies of the isospin triplet 
${^7_\Lambda \rm He}$, ${^7_\Lambda \rm Li}^*$, ${^7_\Lambda \rm Be}$, 
and the $T=1/2$ doublet ${^8_\Lambda \rm Li}$, ${^8_\Lambda \rm Be}$ are investigated within the no-core shell model. 
Calculations are performed based on a hyperon-nucleon potential 
derived from chiral effective field theory at next-to-leading order.
The potential includes the leading charge-symmetry breaking 
(CSB) interaction in the $\Lambda $N channel, whose strength
has been fixed to 
the experimentally known difference of the $\Lambda$ separation 
energies of the mirror hypernuclei ${^4_\Lambda \rm He}$ and ${^4_\Lambda \rm  H}$.
It turns out that the CSB predicted for the $A=7$ systems is
small and agrees with the splittings deduced from the 
empirical binding energies within the experimental uncertainty. 
In case of the $A=8$ doublet, the computed CSB is somewhat larger
than the available experimental value.

\end{abstract}

\maketitle 

\section{Introduction} 
\label{sec:1} 

Charge symmetry breaking (CSB) in $\Lambda$ hypernuclei has been 
experimentally established for many decades. The first  and probably most pronounced evidence  came 
from the difference of the $\La$-separation energies of the 
mirror nuclei ${^4_\Lambda \rm He}$ and ${^4_\Lambda \rm H}$
\cite{Dalitz:1964fu,Raymund:1964an}, eventually followed by
data on other $\Lambda$-hypernuclei isospin multiplets
up to $A=16$ 
\cite{Juric:1973zq,Botta:2016kqd,Achenbach:2016ci,Botta:2019has},
see also \cite{Davis:2005npa,HypernuclearDataBase}. 
However, a solid theoretical understanding of the CSB effects
has been lacking for a long time. Certainly, one of the possible
CSB mechanisms, namely $\La-\Si^0$ mixing, had already been
identified and investigated at an early stage \cite{Dalitz:1964fu}. 
That mechanism facilitates pion exchange between the $\La$ and 
the nucleons, otherwise forbidden by isospin conservation,
and thus yields a long-ranged CSB force. However, with $\La-\Si^0$ mixing
alone, commonly included in elaborate hyperon-nucleon (YN) potentials
like those of the Nijmegen group \cite{Rijken:1999fc},
no quantitative description of the observed CSB 
in the ground ($0^+$) and excited ($1^+$) states of 
${^4_\Lambda \rm He}$-${^4_\Lambda \rm H}$ could be
achieved \cite{Nogga:2001ef}. One could attribute that to the fact 
that the separation-energy difference 
$\Delta B_{\Lambda}(0^+)=B^{0^+}_\Lambda(^4_\Lambda{\rm He})-B^{0^+}_\Lambda(^4_\Lambda{\rm H})$ 
of $340$~keV \cite{Juric:1973zq} and 
$\Delta B_{\Lambda}(1^+) = 240$~keV \cite{Bedjidian:1979ih}
accepted at that time are exceptionally large when 
compared to those found for, say,  the mirror nuclei ${^3\rm H}$ 
and ${^3\rm He}$ of about 80~keV  after the Coulomb-energy correction \cite{Brandenburg:1978ldq}. Indeed, they were also large when compared
to the CSB effects found for heavier $\Lambda$ hypernuclei 
with $A=7$ and $A=8$. In fact,  
cluster model calculations for $A=7-10$ mirror hypernuclei  \cite{Hiyama:2009ki,Hiyama:2012sq,Hiyama:2013owa},
which implemented phenomenological $\La $N  CSB forces that were tuned 
to the splittings found for ${^4_\Lambda \rm He}$-${^4_\Lambda \rm H}$, 
overestimated the CSB splittings for the heavier systems and/or
predicted shifts in the wrong direction. 

In this work, we present a calculation of the binding energies for
the isotriplet 
${^7_\Lambda \rm He}$, ${^7_\Lambda {\rm Li}^*}$, ${^7_\Lambda \rm Be}$
(${^7_\Lambda {\rm Li}^*}$ denotes the excited state of
${^7_\Lambda {\rm Li}}$ with isospin $T=1$),  
as well as of the A=8 doublet  ${^8_\Lambda \rm Li}$, ${^8_\Lambda {\rm Be}}$. 
The study is motivated by the significant experimental and theoretical 
progress that has been made since the last extended calculation by 
Hiyama et al.~\cite{Hiyama:2009ki}. On the experimental side,
there has been a reliable determination of the binding energy
of the ${^7_\Lambda \rm He}$ hypernucleus \cite{Gogami:2016jb}.
Moreover, and more importantly, there has been a re-evaluation of
CSB in the ${^4_\Lambda \rm He}$-${^4_\Lambda \rm H}$ systems. 
Refined data from experiments at J-PARC \cite{Yamamoto:2015avw}
and Mainz \cite{Esser:2015fm,Schulz:2016dt} that became available 
in the years 2015/16 established the splittings to be
$\Delta B_{\Lambda}(0^+) = 233\pm 92$~keV and 
$\Delta B_{\Lambda}(1^+) = -83\pm 94$~keV \cite{Achenbach:2016ci,Gazda:2016ir}. 
Thus, there is a sizable reduction of the CSB effect in the $0^+$
state as compared to the former value. In the $1^+$ state there is 
even a change in the sign, and the new value is practically compatible 
with zero. 

With regard to theory, a consistent description of the charge-symmetry
preserving and CSB components of the $\Lambda N$ interaction has
been achieved within chiral effective field theory (EFT) applying
an appropriate power counting. The resulting potentials yield an
excellent description of the available low-energy $\La p$ and
$\Sigma $N data \cite{Haidenbauer:2013oca,Haidenbauer:2019boi}. 
Earlier studies usually omitted the CSB contact interactions leading to significant 
dependence of the predictions of the CSB for A=4 hypernuclei on details of the interactions \cite{Gazda:2016ir}. This problem could be resolved by taking the CSB contact interactions into 
account and fixing them using  the $A=4$
$\Lambda$-hypernuclei data \cite{Haidenbauer:2021wld}.
In addition, microscopic ``ab initio'' calculations of hypernuclei up to $A=8$
and beyond are feasible now, say, within the no-core shell model (NCSM)
\cite{Wirth:2014ko,Wirth:2017bpw,Wirth:2019cpp,Le:2020zdu}.
As input elementary YN interactions can be used, together with sophisticated  nucleon-nucleon (NN) 
and three-nucleon (3N) forces. Specifically, the important coupling 
between the $\La $N and $\Si $N systems can be fully taken into account and, 
of course, CSB which induces differences in the $\La p$ and
$\La n$ interactions.

The paper is structured in the following way: in 
Sect.~\ref{sec:YN}, we give a brief account of the employed YN interactions. 
Specifically, we explain how the CSB part is determined from 
the separation-energy differences in the $0^+$ and $1^+$ states of 
the $A=4$ hypernuclei ${^4_\Lambda \rm He}$ and ${^4_\Lambda \rm H}$.
In Sect.~\ref{sec:NCSM}, we summarize the treatment of the $A=4-8$ 
hypernuclei within the Jacobi no-core shell model. 
Our results and a discussion of the CSB effects are presented 
in Sect.~\ref{sec:Results}. 
Some further details are relegated to the appendix.
The paper ends with a brief summary. 

\section{Hyperon-nucleon interaction including CSB}
\label{sec:YN}

For the present study, we utilize the YN interactions from
Refs.~\cite{Haidenbauer:2013oca,Haidenbauer:2019boi}, derived 
within SU(3) chiral EFT at next-to-leading
order (NLO). At that order of the chiral expansion, the YN potential 
consists of contributions from one- and two-pseudoscalar-meson
exchange diagrams (involving the Goldstone boson octet $\pi$, $\eta$, $K$)
and from four-baryon contact terms without and with two derivatives.
The two YN interactions are the result of pursuing different 
strategies for fixing the low-energy constants (LECs) that determine 
the strength of the contact interactions.
In the YN interaction from 2013 \cite{Haidenbauer:2013oca}, 
denoted 
by NLO13 in the following, all LECs have been fixed exclusively by a 
fit to the available $\Lambda $N and $\Sigma $N data. The other 
potential \cite{Haidenbauer:2019boi} (NLO19) has been guided 
by the objective to reduce the number of LECs that need to be fixed from 
the YN data by inferring some of them from the NN sector via 
the underlying (though broken) SU(3) flavor symmetry. 
A thorough comparison of the two versions for a range of cutoffs 
can be found in Ref.~\cite{Haidenbauer:2019boi},
where one can see that the two YN interactions yield essentially
equivalent results in the two-body sector. 

The YN potentials NLO13 and NLO19
do not include any explicit CSB contributions.
However, in Ref.~\cite{Haidenbauer:2021wld}, we derived 
the leading CSB interaction within chiral EFT and 
added it to those YN interactions. At the order
considered, CSB contributions arise from a non-zero $\La\La\pi$
coupling constant which is estimated from 
$\Lambda-\Sigma^0$ and $\pi^0-\eta$ mixing, the mass difference between
$K^{\pm}$ and $K^0$, and from two contact terms that
represent short-ranged CSB forces. 
In the actual calculation, the two arising CSB low-energy 
constants (LECs) were 
fixed by considering the known differences in the energy levels
of the $0^+$ and $1^+$ states of the aforementioned
$A=4$ hypernuclei. Then, by construction, the resulting interaction 
describes all low-energy $\Lambda p$ and $\Sigma \mathrm{N}$ scattering 
data, the hypertriton and the CSB in
${^4_\Lambda \rm He}$ and ${^4_\Lambda \rm  H}$ accurately.

For a detailed discussion of the CSB effects we refer the 
reader to \cite{Haidenbauer:2021wld}. As main outcome,  
it turned out that the reproduction of the splittings of
$\Delta B_{\Lambda}(0^+) = 233\pm 92$~keV and 
$\Delta B_{\Lambda}(1^+) = -83\pm 94$~keV \cite{Achenbach:2016ci}
(scenario CSB1 in \cite{Haidenbauer:2021wld})
requires a sizable difference between the strengths of the
$\La p$ and $\La n$ interactions in the $^1S_0$ state,
whereas the modifications in the $^3S_1$ partial wave 
are much smaller. The effects go also in
opposite directions, i.e. while for $^1S_0$ 
the $\La p$ interaction is found to be 
noticeably less attractive than that for $\La n$, 
in case of $^3S_1$ it is slightly more attractive. 
In terms of the difference in the scattering lengths,
$\Delta a^{CSB} = a_{\La p} - a_{\La n}$,
a value of $0.62\pm 0.08$~fm has been predicted for the $^1S_0$ partial wave
and $-0.10\pm 0.02$~fm for the $^3S_1$ \cite{Haidenbauer:2021wld}.

Recently, the STAR collaboration has reported a new measurement for 
the $A=4$ systems, which suggests somewhat different CSB splittings 
of the $0^+$ and $1^+$ states \cite{STAR:2022zrf}. Their results
are $\Delta B_{\Lambda}(0^+) = 160\pm 140({\rm stat}) \pm 100({\rm syst})$~keV 
and 
$\Delta B_{\Lambda}(1^+) = -160\pm 140({\rm stat}) \pm 100({\rm syst})$~keV. 
Of course, considering the sizable statistical and systematic uncertainties,
those values are compatible with the ones cited above, so that 
quantitative conclusions cannot be drawn at present. 
Nevertheless, it is interesting to explore the implication of such 
a possible modification of the CSB in the $A=4$ $\Lambda$ hypernuclei  
for that in the $A=7$ and $8$ systems, though, in view of the 
uncertainties, we 
refrain from doing more elaborate calculations at present. 
Thus, we only re-adjust the two CSB LECs 
for the NLO19 potential in order to reproduce the central values 
of the STAR results. 
We find that the difference in the $\Lambda p$ and $\Lambda n$ scattering lengths 
is somewhat reduced in the $^3S_1$ partial wave,
$\Delta a^{\mathrm{CSB}}(^3S_1) = -0.05$~fm, whereas it slightly increases in 
the $^1S_0$ state, $\Delta a^{\mathrm{CSB}}(^1S_0) = 0.71$~fm. This new set 
of CSB LECs will be referred to as CSB$^*$ and will be employed 
to explore the impact on the splittings in the $A=7$ and $8$ isospin 
multiplets. For a recent and detailed overview on 
the experimental situation regarding the CSB
splittings in the
${^4_\Lambda \rm He}$ and ${^4_\Lambda \rm  H}$
systems see Ref.~\cite{Botta:2019has}.

As shown in previous bound-state calculations, the $\La$ separation
energies of light hypernuclei are not very sensitive to the 
employed NN interaction \cite{Nogga:2001ef,Haidenbauer:2019boi}.
Therefore, we use in all of the calculations presented here the 
same state-of-the-art chiral NN interaction, namely the semi-local momentum-space-regularized (SMS) NN potential of
Ref.~\cite{Reinert:2018ip} at order N$^4$LO$^+$ with cutoff $\La_{\mathrm{N}}=450$~MeV. 
Indeed, the variation of the separation energy for N$^4$LO$^+$ potentials 
with other cutoffs is of the order of $100$~keV for
$^4_{\Lambda}\mathrm{He}/^4_{\Lambda}\mathrm{H}$ \cite{Haidenbauer:2019boi}
and within the range expected from calculations based on phenomenological 
interactions \cite{Nogga:2001ef}. 
A recent dedicated study performed within the NCSM, using however only 
N$^2$LO NN potentials and a LO $YN$ interaction, reported uncertainties of around 
$100$~keV for $A=4$ and of around $400$~keV for A=5 hypernuclei \cite{Gazda:2022fte}. 
Earlier, Wirth and Roth \cite{Wirth:2018ho} found uncertainties of $\approx 200$~keV
and $\approx 400$~keV for $^7_\Lambda$Li and $^9_\Lambda$Be, respectively, 
utilizing N$^3$LO and N$^4$LO NN potentials but also only LO for the $YN$
interaction. 
In preliminary calculations, we observed that the sensitivity of the $\Lambda$ separation energies to the employed NN interactions depends also on the YN interaction itself. 
E.g., for $^3_\Lambda$H, we found variations of the separation energy of 18~keV when using the recent SMS N$^2$LO interactions and 13~keV when using the SMS N$^4$LO$^+$ interactions in
conjunction with NLO19(650). 
With the same NN interactions the variation is 60~keV and 23~keV for the LO(650) YN interaction. The surprisingly large dependence 
of the variation of separation energies on the order of the chiral NN interaction and on the order of the YN interaction might in part explain 
why our estimate of the dependence of the separation energies on the NN interaction is smaller than other values available in the literature. A more detailed study on this issue is in progress but beyond the scope of this work. We also note that Ref.~\cite{Gazda:2022fte} found that the NN force dependence of the CSB in $A=4$ hypernuclei is anyhow smaller due to 
correlations. 

In order to accurately describe the parent nuclei, 
the chiral 3N interaction at order N\textsuperscript{2}LO with the regulator of $\La_{\mathrm{N}} =450$ \cite{LENPIC:2022cyu} is also included. Note 
that such a combination of the NN and 3N forces gives a fairly good description 
for the binding energies of light and medium mass nuclei 
\cite{LENPIC:2022cyu}. It should, however, be stressed that
 although 3N forces contribute moderately to the nuclear and hypernuclear binding energies, their overall effect on the $\Lambda$ separation energies and, in particular, on the CSB
 splittings is expected to be rather small for light  and ground states of $p$-shell hypernuclei \cite{Nuclearandhypernuc:2001wd,Haidenbauer:2019boi,Haidenbauer:2021wld}. The inclusion of 3N forces can improve the description of excited states that are linked to an excited core nucleus. 

\section{Jacobi no-core shell model}
\label{sec:NCSM}
We apply the Jacobi NCSM for calculating the $\Lambda$-separation 
(binding) energies of  the $A=4-8$ hypernuclei. A detailed description of the formalism and of the procedure to extract the binding (separation) energies can be found in Ref.~\cite{Le:2020zdu}. In that reference, and in \cite{Le:2019gjp}, one can also find results
for $^7_\La$Li based on the YN interactions NLO13 and NLO19 without
CSB contribution.
As already mentioned, for the current  study,
we shall employ  chiral NN, 3N and YN potentials
to describe the interactions among the nucleons and between a nucleon and a hyperon, respectively. 
In all calculations, contributions 
   of the NN(YN) potentials in partial waves up to $J = 6(5)$ 
   are included, while for the 3N interaction all partial waves with total angular momentum $ J_{\mathrm{3N}} \leq 9/2$ are taken into account. It has been checked that higher partial waves only contribute negligibly compared to the 
   HO model space uncertainties.  
   In order to speed up the convergence of the NCSM with respect to the model space, all the employed  NN, 3N and YN potentials are 
   SRG-evolved to a flow parameter of $\lambda=1.88$~fm\textsuperscript{-1},  see \cite{Le:2020zdu} and references therein.    The latter is commonly  used in nuclear calculations, which, on the one hand yields rather well-converged nuclear binding energies, and on the other hand, minimizes the possible contribution of SRG-induced 4N and higher-body forces \cite{LENPIC:2022cyu}. 
   Furthermore, in most of the calculations, the SRG-induced 
   YNN interaction with the total angular momentum $J_{\mathrm{YNN}} \leq 5/2$ is also  explicitly included. Based on the contributions of $J_{\mathrm{YNN}} \leq 1/2$, $3/2$ and $5/2$, the contribution from  higher partial waves  $J_{\mathrm{YNN}} \ge 7/2$ is estimated to be negligibly  small and therefore is omitted from the calculations.
   With the proper inclusion of these SRG-induced three-body forces, the 
   otherwise strong dependence of the $\Lambda$ separation energies $B_{\Lambda}$ 
   on the SRG-flow parameter \cite{Wirth:2014ko,Le:2020zdu} is  
   largely removed \cite{Wirth:2018ho,Le:2022jvc}.

It should be further noted that, for large systems like $A=7$ and $8$, 
the extrapolated  NCSM separation energies are afflicted  
with appreciable uncertainties, see \cite{Le:2020zdu} and also Table \ref{tab:A7separation}, which even exceed the experimentally 
found CSB splittings in these systems.
Therefore, it is not advisable to estimate CSB based on the extrapolated 
separation energies. Instead, one can compute the CSB effects directly from 
the corresponding nuclear and hypernuclear energy expectation values for each model space $\mathcal{N}_{max}$
and HO frequency $\omega$. It has been observed that because of the correlations between those binding energies  the directly extracted  $\Delta B_{\Lambda}(\omega, \mathcal{N}_{max})$ converges significantly
faster with respect to $\mathcal{N}_{max}$
and $\omega$ than the individual binding energies and to 
some extent the separation energies, so that a direct comparison with  experiment is possible.
Accordingly, the separation energies difference, say, for $A=4$ systems, can be computed  as   \begin{eqnarray}\label{Eq:pert1}
\Delta B_\Lambda & = &  
B_\Lambda(^4_\Lambda {\rm He}) - B_\Lambda(^4_\Lambda {\rm H}) \cr
& = & 
E(^3{\rm He}) - E(^3{\rm H}) - \big( \, E(^4_\Lambda{\rm He}) -
E(^4_\Lambda{\rm H}) \,  \big).
\end{eqnarray}
Let us further separate contributions from the kinetic energy, and from the NN and
YN interactions  to the total binding energies. This decomposition is justified by the observation that the contributions due to three-body forces are 
negligibly small. 
Hence, the CSB splitting in Eq.~(\ref{Eq:pert1}) can finally be expressed as 
follows
\begin{widetext}
\begin{eqnarray}\label{Eq:pert}
\Delta B_\Lambda & = &  T(^3{\rm He})   - T(^3{\rm H}) - \big (  T(^4_\Lambda{\rm He}) -
T(^4_\Lambda{\rm H}) \big)  +  V_{\mathrm{NN}}(^3{\rm He})   - V_{\mathrm{NN}}(^3{\rm H})\nonumber\\ 
 & \qquad  -&  \big (  V_{\mathrm{NN}}(^4_\Lambda{\rm He}) -
V_{\mathrm{NN}}(^4_\Lambda{\rm H}) \big)  - \big (  V_{\mathrm{YN}}(^4_\Lambda{\rm He}) -
V_{\mathrm{YN}}(^4_\Lambda{\rm H}) \big) \nonumber\\
&  =  &  \Delta T   + \Delta V_{\mathrm{NN}}    + \Delta V_{\mathrm{YN}} \ . 
\end{eqnarray}
\end{widetext}
Note that the operators $V_{NN}$ and $V_{YN}$ employed in Eq.~(\ref{Eq:pert}) 
are also SRG-evolved, like the full Hamiltonian. Furthermore, we will follow the approach in \cite{Haidenbauer:2021wld,NoggaPhD:2001} to estimate the individual
 contributions $\Delta T, \, \Delta V_{NN}$ and $\Delta V_{YN}$ 
 perturbatively based on the two (hyper)nuclear wave functions of
 $^4_\Lambda{\rm He}$ and $^3{\rm He}$  (or $^7_\Lambda{\rm Li^{*}}$ and $^6{\rm Li}$, and 
 $^8_\Lambda{\rm Li}$ and $^7{\rm Li}$ in 
 cases of $A=7$ and 8 systems, respectively). The former  is computed for the YN interactions  that also include the CSB components. Using the wave functions that include CSB effects is strictly speaking a deviation from first order perturbation theory. However, the deviation is of second order and therefore not relevant here. 
 As it has been shown in \cite{Haidenbauer:2021wld,NoggaPhD:2001} and  will be discussed in the following section, such a perturbative estimate of $\Delta B_{\Lambda}$ is a good approximation to the exact calculations. The  wave functions for  $A=4,7$ and 8 hypernuclear ($ A=3,6$ and 7 for nuclear) systems are generated using the largest computationally accessible model spaces, namely $\mathcal{N}_{max} = 24, 10$ and 9, respectively, and at the optimal $\omega_{opt} =16$~MeV that is (or very close to) the variational minimum. 
 
 For estimating numerical uncertainties due to 
 the model space truncation, we have performed the same calculations for two-body interactions 
 at the same $\mathcal{N}_{max}$ and for $\mathcal{N}_{max} + 2$ and with the same $\omega_{opt}$
 and  $\omega_{opt} \pm 2$~MeV. The variation of these calculations 
 gave our uncertainty estimate of 10, 30 and 50~keV for the $A=4$, $7$, and $8$ isospin multiplets, respectively. Note  the larger uncertainty for the $A=8$ doublet because of the smaller accessible model space. 
 
As already said, we will employ the high-order SMS NN interaction 
  with  $\Lambda_{\mathrm{N}} = 450$~MeV
  (SMS N\textsuperscript{4}LO$^+$(450)) \cite{Reinert:2018ip} and the
  N\textsuperscript{2}LO 3N force with the same chiral 
  cutoff \cite{LENPIC:2022cyu}. 
 Two chiral potentials at next-to-leading order, namely  
 NLO13 and NLO19 \cite{Haidenbauer:2013oca,Haidenbauer:2019boi} with a regulator 
 of $\Lambda_{Y}=500$~MeV, are chosen for the YN interaction.  
 We know by experience that the SRG evolution for such low cutoff
 values converges very quickly thanks to the overall small YN potential 
 matrix elements. 
 For larger cutoffs and especially for the NLO13 interaction, which contains sizable off-diagonal potential matrix elements, the ordinary differential equations (ODE) solver used for the SRG evolution demands 
 an extremely small time step for 
 achieving an accurate solution which requires prohibitively large 
 computing resources. 
The predicted difference in the $\Lambda$p and $\Lambda$n scattering 
lengths, i.e.  $\Delta a^{CSB}$, has been found to be basically the same 
for all cutoffs and for the 
 two realizations of the YN interaction \cite{Haidenbauer:2021wld}.
 Obviously, the regulator dependence is efficiently absorbed
 by the contact terms of the CSB component of the YN potentials, 
 when fixing the pertinent LECs from the $A=4$ CSB level splittings. 
 Therefore, we expect that the CSB splittings for $A=7$ and $8$ hypernuclei based on those interactions exhibit likewise a fairly weak 
 or even a negligible cutoff dependence. 
Finally, chiral $Y$NN forces are not considered in the current study,
only those from the SRG evolution. In the following ``3N forces'' 
stands for (the inclusion of) chiral as well as SRG-induced 3N forces, 
unless explicitly stated otherwise.

\section{Results}
\label{sec:Results} 

 \begin{table*}
\renewcommand{\arraystretch}{1.3}

\begin{center}
\begin{tabular}{|l| l | l  | l| l|}
\hline
 \multicolumn{2}{|c|}{}     &    $^4_{\Lambda}\mathrm{H}(0^+)$    &    $^4_{\Lambda}\mathrm{H}(1^+)$ & $^5_{\Lambda}\mathrm{He}$\\
 \hline
\hline
\multirow{3}{*}{\,NLO13-CSB\,}  &  full      &    
$1.551 \pm 0.007$  &   $ 0.823 \pm 0.003$ & $2.22 \pm 0.06$ \\
&    $\lambda=0.765$  &   $1.29 \pm 0.005$   &  $0.77 9\pm 0.02$ & $2.22 \pm 0.04$\\
&    FY   &    $1.513  $  &    $ 0.813$  &\\
\hline
\hline
\multirow{3}{*}{NLO19-CSB}  &  full &    
$1.514 \pm 0.007$  &   $ 1.27 \pm 0.009$ & $3.32 \pm 0.03$\\
&    $\lambda=0.823$  &   $1.41 \pm 0.003$   &  $ 1.131 \pm 0.01$ & $3.35 \pm 0.02$\\
&     FY   & $1.511 $   &   $ 1.268$  & \\
\hline
\hline
\multicolumn{2}{|c|}{experiment}     &     $ 2.16 \pm 0.08$ \cite{Schulz:2016dt}  & $ 1.07 \pm 0.08$ \cite{Schulz:2016dt} & $3.12 \pm 0.02$ \cite{Juric:1973zq}\\
\hline
  \end{tabular}
\end{center}
\caption{\label{tab:Blambda_magic} $\Lambda$-separation energies for the $^4_{\Lambda}\mathrm{H}(0^+, 1^+)$ states and for 
$^5_{\Lambda}\mathrm{He}$,  
computed for the YN potentials NLO13(500) and NLO19(500) including 
the CSB interaction. Listed are our 
full results, with inclusion of the corresponding SRG-induced YNN forces, and those with the YN potentials SRG-evolved to 
the magic flow parameter, $\lambda_{magic} = 0.765$ and $0.823$~fm\textsuperscript{-1}, respectively.  The $B_{\Lambda}$ values are obtained by performing the two-step $\omega$- and $\mathcal{N}$-space  extrapolation, see \cite{Le:2020zdu} for more details. 
The FY calculations are performed with the bare NN, 3N and YN potentials. 
Energies are given in MeV.}
\renewcommand{\arraystretch}{1.0}
\end{table*}
\subsection{Charge symmetry breaking in the  \texorpdfstring{$A=4$}{A=4}  systems} 
\label{sec:A=4}
The hypernuclei $^4_\Lambda$H and $^4_\Lambda$He constitute an
important test case for our calculations, because here we can
directly compare the NCSM predictions with results obtained from
solutions of the Faddeev-Yakubovsky (FY) equations \cite{Haidenbauer:2021wld}.  Table~\ref{tab:Blambda_magic}
shows  the comparison of the separation energies obtained within the 
two methods. 
The  NLO13 and NLO19 potentials with chiral cutoff of $500$~MeV have been employed 
to describe the YN interaction, while  the standard combination of the SMS $\mathrm{N^4LO}^{+}(450)$ NN and $\mathrm{N^2 LO} (450)$ 3N interactions 
is used \cite{LENPIC:2022cyu} for the nucleons. 
For the NCSM calculations, the employed NN, 3N and YN potentials are SRG-evolved to a flow parameter of $\lambda=1.88 \, \mathrm{fm}^{-1}$. Furthermore, the SRG-induced YNN interaction is taken into account so that the separation energies are practically independent of the SRG-flow parameter \cite{Le:2022jvc}. For the FY calculations, the bare NN, 3N and YN interactions have been employed. 
The small discrepancy between 
 the FY results listed in Table~\ref{tab:Blambda_magic} and those provided in \cite{Haidenbauer:2021wld} is essentially due to the contribution of the 3N force, neglected in the latter work, which clearly amounts to less 
 than $50$~keV. 
  It is reassuring to observe that for NLO19 the actual separation energies 
  computed within the NCSM approach agree perfectly with the results of the FY equations, 
  for the ground state as well as for the excited state. 
  The extremely small difference
  could be an indication that the contribution of SRG-induced YNNN forces to 
  the separation energies in the $A=4$ systems are negligibly small. However, 
  for a more quantitative estimate, well-converged calculations using a
  wide range of values for the SRG flow parameter 
 are still necessary. 
 For NLO13, the difference of the FY result and the full calculations is more visible and of the order of $40$~keV indicating larger contributions of the missing SRG-induced YNNN forces in this case which are probably related to the larger $\Lambda$-$\Sigma$ transition matrix elements  \cite{Haidenbauer:2019boi}. We stress that the agreement of the FY and full calculations are still excellent. 
 
 Additionally, the table contains our NCSM results for the
 $\Lambda$ separation energies of $^5_\Lambda$He, which are
 $B_{\Lambda}(^5_\Lambda\mathrm{He}) =2.22 \pm 0.06$~MeV and 
 $3.32 \pm 0.03$~MeV \cite{Le:2022jvc}, respectively.
 Evidently, NLO13 significantly underestimates the $^5_{\Lambda}\mathrm{He}$
 separation energy, while the result for the NLO19 potential is 
 rather close to and only slightly above the experimental value of 
 $B_{\Lambda}(^5_\Lambda\mathrm{He}) =3.12$~MeV. The 
 discrepancy between the two NLO13 and NLO19 predictions signals the
 need for including proper chiral $\La \mathrm{NN}$ 
 and $\Si \mathrm{NN}$ three-body forces \cite{Petschauer:2016ho},
 given that the $\La$p and $\Si$N results of those potentials
 are practically identical. 
 Indeed, three-body forces, with a distinct spin-isospin 
 dependence might also be  needed to 
 bring the $A=4$ results in a better agreement with the experiment.

Finally, we include in Table~\ref{tab:Blambda_magic} results of 
NCSM calculations where only the 
NN potential, SRG-evolved to $\lambda_{\mathrm{NN}}  =1.6$~fm\textsuperscript{-1}, and the two YN potentials, SRG-evolved to the ``magic'' flow parameters, $\lambda_{magic}(\mathrm{NLO19}) = 0.823$~fm\textsuperscript{-1} and  $\lambda_{magic}(\mathrm{NLO13}) = 0.765$~fm\textsuperscript{-1}, 
are employed. The values of $\lambda_{magic}$ for NLO19 and NLO13 
are chosen in such a way that the pertinent full NCSM results 
for the $^5_\Lambda$He separation energy are reproduced.
 Let us remark that our way of fixing $\lambda_{magic}$ 
 here slightly differs from the strategy in \cite{Le:2020zdu} 
 where the experimental value of  $^5_\Lambda\mathrm{He}$ has been used as benchmark. Note that at the SRG parameter of $\lambda_{\mathrm{NN}}  =1.6 $~fm\textsuperscript{-1}, the parent nuclear cores can be fairly well described even when 3N forces are omitted \cite{Le:2020zdu}.
 With $\lambda_{magic}$ fixed to the actual $B_{\Lambda}(^5_\Lambda\mathrm{He})$ for NLO13 and NLO19, we 
 observe a fair to good agreement between the $A=4$ separation energies 
 from the full NCSM calculations and those computed at $\lambda_{magic}$, 
 as can be seen in Table~\ref{tab:Blambda_magic}. 
 The small discrepancy between the two results, up to around $200$~keV 
 for the $0^+$ state and in the order of $100$~keV for $1^+$, can 
 be attributed again to possible contributions from YNN forces 
 \cite{Le:2020zdu,Petschauer:2016ho}.

\begin{table*}
\renewcommand{\arraystretch}{1.2}

\begin{center}
\begin{tabular}{|l|l|r r|  rrr| r  | r|}
\hline
   & $\mathrm{YN}$ & $\Delta T$ & $\Delta V_\mathrm{NN}$ & \multicolumn{3}{c|}{$\Delta V_\mathrm{YN}$}  & $\Delta B_{\Lambda}$ &  $\Delta B_{\Lambda}(FY)$     \\[1pt]
   &    &    &   &  $^1S_0$   & $^3S_1$  &  total &  & \\
\hline
\hline
 \multirow{4}{*}{$(0^+)$} & NLO13          & 17     & -12         &    -3   &   0  & -3      & 3   &   43 \\[1pt]
 & \,NLO13-CSB\,            & 18     & -13      &   152   &  76   &  224    & 229     & 252    \\[1pt]
\cline{2-9}
& NLO19          & 9      & -15       &    -1     &   0   & -1      & -7   &  10  \\[1pt]
& NLO19-CSB            & 9      & -16      &   126     &  118   &  245    & 238  &   238   \\[1pt]
\hline
\hline
  \multirow{4}{*}{$(1^+)$} & NLO13          & 6     & -5           &    0    &   0    & -1     &  0   &   -9 \\[1pt]
& NLO13-CSB            & 6     & -5            &   -114   &  19    & -95   &  -94     & -75   \\[1pt]
\cline{2-9}
&  NLO19          & 5      & -15        &    0        &   0     & -1    & -11   &  5  \\[1pt]
& NLO19-CSB            & 5      & -15      &    -114     &  36    &  -76     & -85  &   -85   \\
\hline
\end{tabular}
\end{center}
\caption{\label{tab:csbA4} Contributions to CSB for $^4_\Lambda$He and $^4_\Lambda$H
in the $0^+$ and $1^+$ states, based on the YN potentials NLO13 
and NLO19 (including 3N forces 
and SRG-induced YNN forces) with cutoff  of
$\Lambda = 500$~MeV. 
The results are for the original potentials 
(without CSB force) and for the scenario CSB1 of Ref.~\cite{Haidenbauer:2021wld}.
FY indicates the exact CSB results extracted from 
Faddeev-Yakubovsky calculations which employ the 
bare  NN, 3N and YN interactions. 
All results are in keV. The estimated uncertainty from the NCSM and FY calculations are $10$ and $20$~keV, respectively. 
The experimental reference values are $\Delta B_{\Lambda}(0^+)=233\pm 92$ and $\Delta B_{\Lambda}(1^+)=-83 \pm 94$~keV. 
}
\renewcommand{\arraystretch}{1.0}
\end{table*}

In Table~\ref{tab:csbA4}, we analyse the CSB  
in the $A=4$ isodoublet $^4_\Lambda$He and $^4_\Lambda$H in detail.
The results are based on NLO13(500) and NLO19(500) as published 
originally \cite{Haidenbauer:2013oca,Haidenbauer:2019boi}
and including a CSB interaction \cite{Haidenbauer:2021wld}
that was adjusted to the experimental splittings
$\Delta B_{\Lambda}(0^+) = 233\pm 92$~keV and $\Delta B_{\Lambda}(1^+) = -83 \pm 94$~keV
(CSB1 of Ref.~\cite{Haidenbauer:2021wld}). 
 Similarly to \cite{Haidenbauer:2021wld}, we break down 
 the different contributions to the total CSB splitting 
 $\Delta B_{\Lambda}$, due to the kinetic energy $\Delta T$, 
 the NN interaction ($\Delta V_\mathrm{NN}$) 
 and the YN interaction ($\Delta V_\mathrm{YN}$), see Eq.~(\ref{Eq:pert}). 
 The perturbatively 
 estimated contributions of the 3N and YNN forces are negligibly small 
 and, therefore, omitted in the table. The CSB contribution $\Delta T$ 
 is also small when using chiral interactions, but contributes with positive sign to the total CSB. The contribution of the nuclear core $\Delta V_\mathrm{NN}$, mostly due to the point Coulomb interaction between the
 protons, is of similar magnitude as $\Delta T$ but comes with a negative sign. 
 As expected, $\Delta V_\mathrm{YN}$ for the original YN potentials is  insignificant. However,
 when the CSB interaction \cite{Haidenbauer:2021wld} is included,  
 $\Delta V_\mathrm{YN}$ becomes sizable and, by construction, the total CSB results for the $0^+$ state as well as for $1^+$ are in line with the 
 aforementioned empirical information.
 
 Also for the CSB splittings, we can compare our NCSM results with those
 obtained by solving the FY equations, cf. the last column in 
 Table~\ref{tab:csbA4}. Again, there is good agreement between 
 the two calculations within the estimated uncertainties. 
 Note that the FY values are from an exact solution of the equations.
 The comparison of perturbative and exact CSB results in
 Tables~6 and 7 of Ref.~\cite{Haidenbauer:2021wld} reveals that there
 is very little difference. In addition, 
 for $A=7$ systems, we have also explicitly studied and observed a discrepancy of only less than ten keV  between the perturbativly estimated CSB and the CSB results that are computed based on the expectation values of the  $T$, $V_{NN}$ and $V_{YN}$ operators estimated with respect to the corresponding hyper(nuclear) wavefunctions $^7_{\Lambda}\mathrm{Be} (^6\mathrm{Be})$ and  $^7_{\Lambda}\mathrm{Li^{*}} (^6\mathrm{Li})$. Let us again stress that
 due to the large  uncertainties of the extrapolated separation energies  for the $A \ge 7$ systems, see Table~\ref{tab:A7separation}, a direct extraction of CSB splittings based on those separation energies is not useful. One could also calculate the CSB differences 
 for each model space separately and study the model space 
 and $\omega$ dependence more carefully. For $A=4$ and $A=7$ hypernuclei,
 our results for this approach were also consistent with the perturbative estimate, but there was still a visible dependence on the model space 
 size and HO frequencies which made the extraction of an uncertainty rather difficult. 
 We therefore favor the perturbative approach which is robust and computationally less demanding and use it for obtaining the 
 CSB effects in the
 $A=7$ and $8$ $\Lambda$ hypernuclei below. 
 
 Let us now  have a closer look at the different contributions of the $^1S_0$ and $^3S_1$ partial waves, $\Delta V_{\mathrm{YN}(^1S_0)}$ and 
 $\Delta V_{\mathrm{YN}(^3S_1)}$, to the total 
 $\Delta V_\mathrm{YN}$. From the fourth column in Table~\ref{tab:csbA4}, it follows that
 $\Delta V_{\mathrm{YN}(^1S_0)}$ and $\Delta V_{\mathrm{YN}(^3S_1)}$ are sizable and of 
 the same sign in the $0^+$ state, resulting in a 
 large $\Delta V_{\mathrm{YN}(0^+)}$. 
 In the excited state, the two contributions are, however, smaller and of opposite sign so that there is some cancellation. 
 The signs of the two contributions $\Delta V_{\mathrm{YN}(^1S_0)}$ and 
 $\Delta V_{\mathrm{YN}(^3S_1)}$ 
 are directly related to the different strengths of the $\Lambda n$ 
 and $\Lambda p$ interactions in the singlet and triplet states,  
 as manifested by the respective scattering lengths, and to the 
 relative weights
 of the $\Lambda n$ and $\Lambda p$ components in those spin states. 
 More details are given in the appendix.

\begin{table*}
\renewcommand{\arraystretch}{1.3}
\begin{center}
\begin{tabular}{|l| l l |ll|ll|}
\hline
 &   \multicolumn{2}{c|}{NLO19}  &  \multicolumn{2}{c|}{NLO13}   &   
 \multicolumn{2}{c|}{experiment}  \\
  & full  & $\lambda=0.823$  & full  & $\lambda=0.765$ &  &  \\
\hline
  $^7_\La$Be   & $5.54 \pm 0.22$ & $5.44 \pm 0.03$  & $4.30   \pm 0.47$  &$ 4.53 \pm0.34$ &$5.16 \pm 0.08$ &  \\
  $^7_\La$Li$^*$   & $5.64 \pm 0.28$ &  $5.49 \pm 0.04$ & $4.42   \pm  0.58$  & $4.59 \pm 0.34$ & $5.26 \pm 0.03$
  &  $5.53\pm 0.13$   \\
  $^7_\La$He & $5.64 \pm 0.27$  & $5.43 \pm 0.06$  & $4.39 \pm 0.54$  & $4.45 \pm 0.35 $& & $5.55 \pm 0.1\phantom{0}$ \\
\hline
\hline
  $^8_\La$Be   &  &$ 7.15 \pm0.10$ & & $5.56 \pm 0.25$  & $6.84 \pm 0.05$ &  \\
    $^8_\La$Li & $7.33\pm 1.15$  
    &$ 7.17 \pm0.10$  & $5.75\pm 1.08$ & $5.57 \pm 0.30$  &  
    $6.80 \pm 0.03$ &  \\
  \hline
\end{tabular}
\end{center}
\caption{\label{tab:A7separation} $\Lambda$ separation energies for the
$A = 7$ and $8$ systems, computed for NLO13(500) and NLO19(500) including
the SRG-induced YNN forces (full), and at the magic flow parameters 
(third and fifth columns). 
Note that the separation energies of  $A=7 (8)$ for   NLO19 at $\lambda=0.823$~fm\textsuperscript{-1} have been computed with model spaces up to $\mathcal{N}_{max}=12(11)$, whereas the other calculations are performed with $\mathcal{N}_{max}=10 (9)$. The  listed $B_{\Lambda}$ values are obtained by performing the two-step $\omega$- and $\mathcal{N}$-space  extrapolation, see \cite{Le:2020zdu} for more details.  Values from emulsion (left) and counter (right) experiments 
are taken from the compilation in Ref.~\cite{Botta:2016kqd}. Energies are given in MeV.
}
\renewcommand{\arraystretch}{1.0}
\end{table*}

\subsection{Charge symmetry breaking in the \texorpdfstring{$A=7$}{A=7} and 8 systems}
\label{sec:A=7}
We now employ the NLO13(500) and NLO19(500) potentials to study CSB 
in the $A=7$ isotriplet and the $A=8$ isodoublet. 
Predictions for the separation energies of the $(1/2^+,1)$ mirror hypernuclei  
$^7_\La$He, $^7_\La$Li$^*$, $^7_\La$Be without CSB terms are provided in Table~\ref{tab:A7separation}. The values listed in the second and fourth columns 
have been obtained with inclusion of both the chiral and SRG-induced 3N forces as well 
as of the SRG-induced YNN interactions, whereas $B_{\Lambda}$ displayed in the 
third and fifth columns is computed at the corresponding $\lambda_{magic}$. 
Obviously, there is a fairly good agreement between the separation energies 
extracted from the full calculations and the one at $\lambda_{magic}$.  
This confirms our observation in \cite{Le:2019gjp,Le:2020zdu} 
that the magic SRG-flow parameters can be utilized to simplify the
calculation of light hypernuclear systems.
The NLO13 interaction predicts separation energies of $B_{\Lambda} = 4.30 \pm 0.47$,  $4.42 \pm 0.58$  and $4.39 \pm 0.54$~MeV for $^7_{\Lambda}\mathrm{Be}$, $^7_{\Lambda}\mathrm{Li^{*}}$, and $^7_{\Lambda}\mathrm{He}$, respectively, 
and, thus, underestimates the empirical values by about $1$~MeV. On the other 
hand, the results based on NLO19 are rather close to  experiment. 
In particular, the obtained separation energies for the $T_3=0 $ and 
$T_3 = -1$ members 
$B_{\Lambda}(^7_{\Lambda}\mathrm{Li^{*}} ) = 5.64 \pm 0.28$~MeV and 
$B_{\Lambda}(^7_{\Lambda}\mathrm{He} ) = 5.64 \pm 0.27$~MeV  
are perfectly in line with the values of 
$B_{\Lambda}(^7_{\Lambda}\mathrm{Li^{*}} )= 5.53 \pm 0.13$ and $B_{\Lambda}(^7_{\Lambda}\mathrm{He} )=5.55\pm 0.13$~MeV, extracted 
from counter experiments with an absolute energy
calibration \cite{Botta:2016kqd}. 
For the $^7_{\Lambda}\mathrm{Be}$ hypernucleus, we obtain a separation 
energy of $B_{\Lambda}( ^7_{\Lambda}\mathrm{Be})= 5.54 \pm 0.22$~MeV, which exceeds 
the emulsion result of $B_{\Lambda}(^7_{\Lambda}\mathrm{Be}) = 5.16 \pm 0.08$~MeV\cite{Botta:2016kqd}. However, considering the unresolved  
difference of $270\pm 170$~keV between the 
$B_{\Lambda}(^7_{\Lambda}\mathrm{Li^{*}})$ determinations in 
counter and emulsion experiments, cf. Table~\ref{tab:A7separation}, 
the actual discrepancy for $B_{\Lambda}(^7_{\Lambda}\mathrm{Be})$
could be much smaller. Hopefully, future counter experiments 
will settle this issue.

The separation energies for the $A=8$ systems are likewise summarized in
Table~\ref{tab:A7separation}. The results for $^8_{\Lambda}\mathrm{Li}$
with both 3N forces and SRG-induced YNN interactions included
are obtained from the full calculations with model space up 
to $\mathcal{N}_{max}=9$. Extending the calculation for model spaces up 
to $\mathcal{N}_{max}=11$ will definitely help to reduce the estimated 
errors. Unfortunately, such a calculation is
very CPU-time consuming and we need to postpone it to a future study.
Nevertheless, in spite of the large uncertainty, it clearly follows from Table~\ref{tab:A7separation} that the separation energy for 
$^8_{\Lambda}\mathrm{Li}$ for the NLO13 potential is substantially too 
low whereas the prediction for NLO19,
$B_{\Lambda}(^8_{\Lambda}\mathrm{Li}) = 7.33 \pm 1.15$~MeV, exceeds the 
empirical value of $B_{\Lambda}= 6.80 \pm 0.03$~MeV
\cite{Botta:2016kqd} only moderately.  Again, 
 the difference in the predictions of 
NLO13 and NLO19 can be attributed to possible contributions of 
chiral YNN forces \cite{Haidenbauer:2019boi,Le:2020zdu,Petschauer:2016ho}. 
Although full calculations for $^8_{\Lambda}\mathrm{Be}$ have not 
 been performed yet, a result for $B_{\Lambda}(^8_{\Lambda}\mathrm{Be})$ 
 very similar to that for the  $^8_{\Lambda}\mathrm{Li}$ hypernucleus 
 can be expected. 
 $B_{\Lambda}$ values for $^8_{\Lambda}\mathrm{Be}$ and  $^8_{\Lambda}\mathrm{Li}$ computed at the magic SRG-flow parameters are
 given in the third and fifth columns of Table~\ref{tab:A7separation}. Evidently, the obtained separation energies, e.g. $B_{\Lambda}(^8_{\Lambda}\mathrm{Li}, \lambda_{magic}) = 7.17 \pm 0.10$~MeV, is close to the result of $B_{\Lambda}(^8_{\Lambda}\mathrm{Li}) = 7.33 \pm 1.15$~MeV of the full calculations. This is not too
 surprising in view of what we had already
 observed in the pertinent comparison for the $A=4$ and $7$ systems. 
 Note that $B_{\Lambda}(^8_{\Lambda}\mathrm{Li})$ based on 
 $\lambda_{\rm magic}$ exceeds the value from the
 emulsion experiment only by $0.37 \pm 0.13$~MeV.
 Anyway, in view of the rather good agreement of 
 our predictions for the $A=7$ systems with the separation energies from counter experiments, corresponding measurements for $A=8$ hypernuclei, 
 that could either confirm or revise the emulsion results, are 
 desirable.

\begin{table*}
\renewcommand{\arraystretch}{1.5}
\begin{center}
\begin{tabular}{|l|l|r r|  rrr| r  | }
\hline
\multicolumn{2}{|c|}{} & $\Delta T$ & $\Delta V_\mathrm{NN}$ & \multicolumn{3}{c|}{$\Delta V_\mathrm{YN}$}  & $\Delta B_{\Lambda}$       \\[1pt]
\multicolumn{2}{|c|}{}  &  &  &  $^1S_0$  & $^3S_1$  &  total &  \\
\hline
\hline
 \multirow{4}{*}{$^7_\La$Be-$^7_\La$Li$^*$} & NLO13 &7     & -24          & -1  & 0  & 0             & -17      \\
&  NLO13-CSB & 8    & -24              & -49 & 26   & -24        & -40    \\
\cline{2-8}
&  NLO19 &6     & -40              &-1  & 0        &0     & -34  \\
&  NLO19-CSB &6     & -41              & -43 & 42   & 9        & -35      \\
\cline{2-8}
& Hiyama~\cite{Hiyama:2009ki}  &      &   -70 &    &  &  200 & 150\\
& Gal~\cite{Gal:2015bfa} &  3   & -70  &  &  & 50  &  -17\\
\cline{2-8}
& \,experiment~\cite{Botta:2019has}\, &  &  &  &  &   & $-100 \pm 90 $\\
\hline
\hline
 \multirow{4}{*}{$^7_\La$Li$^*$-$^7_\La$He} & NLO13 & 8    & -13                & 0 & 0    & 0     & -5         \\
 & NLO13-CSB &7     & -14               & -49 & 26    & -24         & -31        \\
\cline{2-8}
 & NLO19 & 5     & -22                 &-43  & 42   & 0         & -17         \\
 & NLO19-CSB & 5     & -21                 & -38 &  37    & -1      & -16        \\
\cline{2-8}
& Hiyama~\cite{Hiyama:2009ki}  &  & -80   &    &  & 200 & 130\\
& Gal~\cite{Gal:HYP2015} &  2   & -80  &  &  & 50  &  -28\\
\cline{2-8}
& experiment~\cite{Botta:2019has} &  &  &  &  &  & $-20 \pm 230 $\footnote{The difference between
B$_\Lambda$($^7_\Lambda$Li$^*$) and B$_\Lambda$($^7_\Lambda$He) 
is $-20\pm 230$~keV for the FINUDA and JLab results, but $-50\pm 190$~keV when the 
revised SKS and JLab results are used \cite{Botta:2019has}.}\\
& & & & & & & $-50\pm 190$$\phantom{^a}$ \\
\hline
\hline
 \multirow{4}{*}{$^8_\La$Be-$^8_\La$Li}
& NLO13          & 12     & 8         &    -2   &   0  & -4      & 16     \\
& NLO13-CSB            & 12     & 7     &   100   &  56    &  159    &  178          \\
\cline{2-8}
& NLO19          & 7      & -11      &    -1     &   0     & -2      & -6     \\
& NLO19-CSB            & 6      & -11       &   62     &  79  &  147    & 143      \\
\cline{2-8}
& Hiyama~\cite{Hiyama:2009ki}  &      & 40   &    &  & & 160\\
& Gal~\cite{Gal:2015bfa}  & 11    & -81  &  & &  119  & 49\\
\cline{2-8}
& experiment \cite{Botta:2016kqd}  &   &    &   &  &  & $40 \pm 60$ \\
\hline
\end{tabular}
\end{center}
\caption{\label{tab:A7_csb} 
Contributions to CSB in the $A=7$ and $8$ isospin multiplets, 
based on the YN potentials NLO13(500) and NLO19(500) (including 
3N forces and SRG-induced YNN interactions). The results are for the original potentials 
(without CSB force) and for the scenario CSB1, see text. 
Results by Gal \cite{Gal:2015bfa} and by Hiyama et al.~\cite{Hiyama:2009ki} 
are included for the ease of comparison. 
All energies are in keV. The estimated uncertainties for $A=7$ and $8$ systems are  30 and 50~keV, respectively.  
}
\renewcommand{\arraystretch}{1.0}
\end{table*}

Table~\ref{tab:A7_csb} provides a detailed view on the CSB splittings 
for the three members of the $A=7$ isotriplet, by comparing 
$^7_{\Lambda}\mathrm{Be}$-$^7_{\Lambda}\mathrm{Li}^*$ and  $^7_{\Lambda}\mathrm{Li}^*$-$^7_{\Lambda}\mathrm{He}$,  
computed for NLO13 and NLO19 without and with CSB interaction. 
The 3N forces and the SRG-induced YNN interactions are explicitly taken into account. 
One sees that, despite the substantial discrepancy in the predicted $\Lambda$ 
separation energies, 
the two potentials yield comparable CSB results in the $A=7$ systems. 
The overall CSB effect is rather small, 
with as well as without the CSB part of the potentials,
and consistent with the experiment, both in magnitude and sign.
It is also interesting to note that, 
 like in the $1^+$ state of the $A=4$ systems, the $^1S_0$ and $^3S_1$ 
 states contribute with opposite signs to the total 
 $\Delta V_{\mathrm{YN}}$, which, in turn, leads to a small total CSB for the $A=7$ isotriplet. 

We include also results of former studies for the ease of comparison. 
Those of Gal~\cite{Gal:2015bfa,Gal:HYP2015} were computed by employing 
a shell-model approach in combination with an effective $\La\Si$ 
coupling model. 
The $A=7$ calculation by Hiyama et al.~\cite{Hiyama:2009ki} is done 
within a ($\Lambda + \mathrm{N + N} + \alpha$) four-body cluster model. 
Surprisingly, our prediction for 
$^7_{\Lambda}\mathrm{Be}$-$^7_{\Lambda}\mathrm{Li^{*}}$ for the
original NLO13 potential (without CSB interaction), 
$\Delta B_{\Lambda}(\mathrm{NLO13}) = -17$~keV,
is  identical to the CSB estimated by Gal~\cite{Gal:2015bfa}. However, the individual contributions
$\Delta T$, $\Delta V_\mathrm{NN}$, and $\Delta V_\mathrm{YN}$ differ substantially. 
For example, the NLO13 potential yields a vanishing 
$\Delta V_\mathrm{YN}$ (because, as said, there is no CSB part), 
whereas in Gal's calculation this contribution amounts 
to $50$~keV. $\Delta V_\mathrm{YN}$ evaluated for the actual chiral 
CSB interaction is of opposite sign and smaller.
There is also a large difference in $\Delta V_\mathrm{NN}$ (that quantity 
includes also the Coulomb effect). 
Note that $\Delta V_\mathrm{NN}$ used by Gal is taken from the cluster-model 
study of Hiyama et al.~\cite{Hiyama:2009ki} whereas our value is 
calculated consistently within the NCSM. 

CSB results for the two $A=8$ mirror nuclei are 
listed at the lower end of Table~\ref{tab:A7_csb}.  
When using the potentials NLO13 and NLO19 without the CSB 
interaction, a negligibly small CSB is predicted 
 for $^8_{\Lambda}\mathrm{Be}$-$^8_{\Lambda}\mathrm{Li}$, namely 
 $\Delta B_{\Lambda} = 16 \pm 50$~keV and $-6 \pm 50$~keV, respectively. This is, however, well in line with the empirical CSB of $40 \pm 60$~keV \cite{Botta:2016kqd} based on the separation energies determined in emulsion experiments. A similarly small $ \Delta B_{\Lambda}$ was also predicted  by Gal in \cite{Gal:2015bfa}, but, in contrast to the rather small $\Delta V_\mathrm{NN}$ contribution in our calculation, e.g. $\Delta V_\mathrm{NN} =-11$~keV
 for NLO19, Gal assigned a significantly larger value to  $\Delta V_\mathrm{NN}$, namely $\Delta V_\mathrm{NN} = -81$~keV. The latter was not computed directly but taken from the shell model calculation by Millener \cite{Gal:2015bfa}.

 With the CSB interaction included, both the NLO13 and NLO19 potentials yield rather sizable CSB results, 
 $\Delta B_{\Lambda}(\mathrm{NLO13}) =177 \pm 50$~keV and  
 $\Delta B_{\Lambda}(\mathrm{NLO19}) =143 \pm 50$~keV. In this case, the $^1S_0$ and $^3S_1$ partial-wave contributions are large, and more importantly, are of the same sign, and, therefore, add up to a 
 pronounced total CSB. This exactly resembles the situation 
 for the $0^+$ states of the $A=4$ mirror hypernuclei discussed in Sect.~\ref{sec:A=4}. Indeed, it is conceivable that a fairly 
 large splitting in the $0^+$ state, as presently established, 
 implies automatically a likewise significant CSB splitting in 
 ${^8_{\Lambda}\mathrm{Be}}$--${^8_{\Lambda}\mathrm{Li}}$.
 Interestingly, the predictions of NLO13 and NLO19 with CSB interaction are comparable to the value of 
 $\Delta B_{\Lambda} = 160 $~keV obtained in a 
 ($\Lambda$+$\alpha$+$^3$He/$t$) three-body cluster calculation  
 by Hiyama et al.~\cite{Hiyama:2009ki,Hiyama:2002yj}. 
 However, it should be noted that the
 phenomenological CSB YN interaction used in Ref.~\cite{Hiyama:2009ki} 
 was fitted to 
 an outdated CSB splitting in the $A=4 $ systems, namely  
 $\Delta B_{\Lambda}(0^+) = 350 \pm 60$~keV and $\Delta B_{\Lambda}(1^+) = 240 \pm 60$~keV. 
 Also, it should be said that, when using only the charge symmetric
 phenomenological interactions adjusted so that the experimental value 
 of $B_{\Lambda}(^8_{\Lambda}\mathrm{Li}) = 6.80$~MeV is reproduced, 
 Hiyama et al. obtained a separation energy of
 $B_{\Lambda} = 6.72$~MeV for $^8_{\Lambda}\mathrm{Be}$. 
 The difference of $-80$~keV between $B_{\Lambda}(^8_{\Lambda}\mathrm{Be})$ and  $B_{\Lambda}(^8_{\Lambda}\mathrm{Li})$ was then attributed to the difference of the Coulomb interaction \cite{Hiyama:2009ki}, which only
 amounts to about $10$~keV in our calculations.  
 
 \begin{table*}
\renewcommand{\arraystretch}{1.3}
\begin{center}
\begin{tabular}{|l| c|c c |  c  c | c|}
\hline
     &  & \multicolumn{2}{c|}{${^4_{\Lambda}\mathrm{He}} - {^4_{\Lambda}\mathrm{H}}$}  &  ${^7_{\Lambda}\mathrm{Be}} - {^7_{\Lambda}\mathrm{Li^{*}}}$  &  ${^7_{\Lambda}\mathrm{Li^{*}}}- {^7_{\Lambda}\mathrm{He}}$ & ${^8_{\Lambda}\mathrm{Be}} - {^8_{\Lambda}\mathrm{Li}}$    \\[3pt]
    &   &  $ {0^+} $    &     $ {1^+} $  &   &  & \\   
\hline
\hline
NLO13-CSB  &  full   & 229  &-94  & -40  & -5  & 178     \\
NLO13-CSB & $\lambda = 0.765$   & 213  &-80  & -10  & 0  & 204     \\
\hline
\hline
NLO19-CSB  &  full  & 238  &-85  & -35  & -16  & 143     \\
NLO19-CSB &$\lambda = 0.823$   &  210     &   -71  &-26  & -3 &135  \\
\,NLO19-CSB$^*$\, & $\lambda = 0.823$ & 130 &  -135  &-83  & -62 & 74  \\
\hline
\end{tabular}
\end{center}
\caption{\label{tab:csb0_magic} CSB splittings $\Delta B_\Lambda$ 
(in keV) 
for $A=4-8$ systems. Results for the full calculation and 
those based on the YN potentials NLO13 and NLO19, 
SRG-evolved to $\lambda_{magic}=0.765 $ and
$\lambda_{magic}=0.823 $~fm\textsuperscript{-1}, respectively,
are compared. 
CSB$^*$ corresponds to a CSB interaction adjusted to the 
new STAR data \cite{STAR:2022zrf}, see text. The estimated uncertainties for $A=4,7$ and 8 systems are 10, 30 and 50~keV, respectively.
}
\renewcommand{\arraystretch}{1.0}
\end{table*}

 Finally, for illustration, we compare in Table~\ref{tab:csb0_magic}
 CSB results based on calculations with the magic flow parameter $\lambda_{magic}$, i.e. of calculations without 3N forces and without 
 the SRG-induced YNN interaction, with the full results. 
Furthermore, we discuss the implications of a somewhat different 
CSB splitting
in the $A=4$ system, as suggested by a recent STAR measurement~\cite{STAR:2022zrf}.
For the latter aspect a new scenario is introduced, called CSB$^*$, 
where the LECs of the CSB interaction have been re-adjusted to match 
the CSB splittings reported by STAR, namely $160 \pm 160$~keV for 
the $0^+$ state and $-160 \pm 140$~keV for $1^+$.

It is reassuring though not surprising that the full $A=4$ 
CSB results differ from the values computed at $\lambda_{magic}$ 
by at most $30$~keV. 
The CSB splittings for the $A=7$ systems, computed at the magic 
SRG-flow parameters, are likewise in rather good agreement with 
the results extracted from the full calculations.
The same is also true for the $A=8$ isodoublet. 
 Apparently, the magic SRG-flow parameter is a fairly reliable starting point for 
 studying the separation energies as well as CSB effects in light hypernuclei.
 This important observation could help to significantly save computational resources.
 
Regarding the new STAR data, it clearly sticks out from Table~\ref{tab:csb0_magic} that the corresponding scenario 
CSB$^*$ yields somewhat larger CSB for 
$^7_{\Lambda}\mathrm{Be}$-$^7_{\Lambda}\mathrm{Li}^*$  and  $^7_{\Lambda}\mathrm{Li}^*$-$^7_{\Lambda}\mathrm{He}$, 
$\Delta B_{\Lambda} = -83 \pm 30$~keV and 
$\Delta B_{\Lambda} = -62 \pm 30$~keV, respectively, as compared to the values of  $\Delta B_{\Lambda} = -26 \pm 30$~keV and  
$\Delta B_{\Lambda}=-3 \pm 30 $~keV, predicted by the standard CSB interaction. However, overall, both the CSB$^*$ 
and CSB results are still consistent with the experimental values of  $\Delta B_{\Lambda}(^7_{\Lambda}\mathrm{Be}$-$^7_{\Lambda}\mathrm{Li^{*}}) = -100\pm 90$~keV and 
$\Delta B_{\Lambda}(^7_{\Lambda}\mathrm{Li}^*$-$^7_{\Lambda}\mathrm{He}) = -20\pm 230$~keV \cite{Botta:2019has}.
 Also in case of the A=8 isodoublet the splitting 
 of $74 \pm 50$~keV predicted for the scenario CSB$^*$ 
 is well in line with the experimental value of $40 \pm 60$~keV \cite{Botta:2016kqd}.

\section{Summary}

In this work, we have presented results for the separation energies 
of the isospin triplet ${^7_\Lambda \rm He}$, ${^7_\Lambda \rm Li}^*$, 
${^7_\Lambda \rm Be}$, and the $T=1/2$ doublet 
${^8_\Lambda \rm Li}$, ${^8_\Lambda \rm Be}$, calculated within the 
NCSM. The underlying YN interactions,
taken from Refs.~\cite{Haidenbauer:2013oca,Haidenbauer:2019boi,Haidenbauer:2021wld},
are derived from chiral effective field theory at NLO.
The potentials include the leading CSB 
interaction in the $\Lambda \mathrm{N}$ channel, whose strength
has been fixed to the experimental difference of the $\Lambda$ separation 
energies of the mirror hypernuclei ${^4_\Lambda \rm He}$ and 
${^4_\Lambda \rm  H}$ as established by the J-PARC and Mainz data
\cite{Yamamoto:2015avw,Esser:2015fm,Schulz:2016dt}.
In order to speed up the convergence of the NCSM with respect 
to the model space, all included interactions are SRG-evolved and 
the arising SRG-induced three-body forces are taken into account. 

We have found that the YN potential NLO13 \cite{Haidenbauer:2013oca} 
produces too low separation energies for the $A=7$ and $8$ systems 
considered in the present work. However, the predictions for the YN potential 
from 2019 (NLO19) \cite{Haidenbauer:2019boi} agree quite well with the
experimental values for ${^7_\Lambda \rm He}$ and ${^7_\Lambda \rm Li}^*$,
deduced from counter experiments. On the other hand, separation energies 
obtained from emulsion experiments for ${^7_\Lambda \rm Be}$ and for 
the $A=8$
hypernuclei ${^8_\Lambda \rm Be}$ and ${^8_\Lambda \rm Li}$ are overestimated. 
For either potentials 
the discrepancies between theory and experiments and the differences of NLO13 and NLO19 might be a signal for  the necessity of chiral $\Lambda $NN  and $\Sigma $NN three-body forces \cite{Petschauer:2016ho}, which have been not included so far in our
calculations. 
At the same time, one has to keep in mind that the experimental situation for the hypernuclei studied in the present work is not 
yet settled, specifically concerning the emulsion data, see the discussion in
Refs.~\cite{Botta:2016kqd,Achenbach:2016ci,Botta:2019has}. 

With regard to CSB, the predicted values for 
the $A=7$ systems are small and agree with the splittings deduced 
from the empirical binding energies within the experimental uncertainty. 
In case of the $A=8$ doublet, the computed CSB is somewhat larger
than the available experimental value. We stress that possible 
YNN three-body forces should have only a minor influence on the 
calculated CSB splittings. In view of the still uncertain experimental situation, we also considered a scenario motivated by recent data from the STAR collaboration for $A=4$. We found slightly increased values for the CSB in $A=7$ and a significant reduction in $A=8$. The different effects in $A=7$ and 8 hypernuclei are related to contributions of different sign in the $^1$S$_0$ and $^3$S$_1$ partial waves. Accurate experimental data in these systems will therefore allow one to independently check the CSB deduced from $A=4$ hypernuclei.

We have also explored in  detail the possibility to use the so-called magic flow 
parameter of the SRG evolution in the actual NCSM computations. 
In this case, in contrast to the full calculation which includes 3N 
forces and the SRG-induced YNN interaction, only two-body interactions are
taken into account. In such a scenario, one can save a significant amount of 
computational resources. But then, as a consequence,  the results depend on the 
actual value of the SRG-flow parameter. We consider the option to fix its
value by requiring that the same $^5_\Lambda$He separation energies are 
obtained as in the full NCSM calculation. It turned out that the 
separation energies obtained with that choice of the flow parameter  are fairly close to the
full results. This suggests that the ``magic'' SRG-flow parameter is a 
fairly reliable starting point for studying the separation energies as 
well as CSB effects in light hypernuclei in an ``inexpensive'' way.

\vskip 0.4cm
\section*{Acknowledgements}
We thank Stefan Petschauer for his collaboration in the early stage of
this work. 
This project is part of the ERC Advanced Grant ``EXOTIC'' supported 
the European Research Council (ERC) under the European Union's Horizon 2020 
research and innovation programme (grant agreement No. 101018170).
This work is further supported in part by the DFG and the NSFC through
funds provided to the Sino-German CRC 110 ``Symmetries and
the Emergence of Structure in QCD'' (DFG Project-ID 196253076 - TRR 110), the VolkswagenStiftung (grant no. 93562), and by the MKW NRW under the funding code NW21-024-A.
The work of UGM was supported in part by The Chinese Academy
of Sciences (CAS) President's International Fellowship Initiative (PIFI)
(grant no.~2018DM0034). We also acknowledge support of the THEIA net-working activity 
of the Strong 2020 Project. The numerical calculations were performed on JURECA
and the JURECA-Booster of the J\"ulich Supercomputing Centre, J\"ulich, Germany.

\appendix

\section{Contribution of \texorpdfstring{$^1S_0$}{1S0} and \texorpdfstring{$^3S_1$}{3S0} partial waves to \texorpdfstring{$\Delta V_{YN}$}{Delta VYN}}

\begin{table}
\renewcommand{\arraystretch}{1.3}
\begin{center}
\begin{tabular}{| l |  r r |  r  r | r r |}
\hline
&         \multicolumn{2}{c|}{$^1S_0$}   &    \multicolumn{2}{c|}{$^3S_1$}  &    \multicolumn{2}{c|}{$\langle V_{\mathrm{YN}} \rangle$} \\
&    $\Lambda p $   &   $\Lambda n $  & $\Lambda p $      &  $\Lambda n $     & $^1S_0$    &  $^3S_1$\\
\hline
\hline
$^4_{\Lambda}\mathrm{He}(0^+)$  &  13.92   & 27.60   & 44.54  &  0.42   & -4.383   &  -3.916\\
$^4_{\Lambda}\mathrm{H}(0^+)$  &  27.1   & 13.66   & 0.41  &  43.79   & -4.257   &  -3.797\\
\hline
\hline
$^4_{\Lambda}\mathrm{He}(1^+)$  &  14.48   & 0.13   & 42.47  &  27.07   & -1.383   &  -5.743\\
$^4_{\Lambda}\mathrm{H}(1^+)$    & 0.128  &14.48   & 27.16  & 42.48  &-1.423  & -5.8685\\
\hline
\hline
$^7_{\Lambda}\mathrm{Be}$  &  11.13   & 7.22   & 33.25  &  21.67   & -3.733  &  -9.364\\
$^7_{\Lambda}\mathrm{Li^{*}}$  &  9.17   & 9.17   & 27.44  &  27.44   & -3.768  &  -9.321\\
$^7_{\Lambda}\mathrm{He}$  & 7.22    & 11.10  &  21.65 &  33.13   &-3.802   &-9.278  \\
\hline
\hline
$^8_{\Lambda}\mathrm{Be}$  &  9.49   &12.24   & 28.67  &  19.33   & -5.315  &  -9.959\\
$^8_{\Lambda}\mathrm{Li}$  & 11.71     &  9.5 &  19.84 &   28.58  & -5.254   & -9.876 \\
\hline
\end{tabular}
\end{center}
\caption{\label{tab:Prob} Probability (in $\%$) of finding $\Lambda p$
 and $\Lambda n$ pairs in the $A=4\,-\,8$ wave functions, and 
 the contributions of the $^1S_0$ and $^3S_1$  $\Lambda \mathrm{N}$ partial waves to the 
 expectation value  $\langle V_{\mathrm{YN}}\rangle$ (in MeV). 
 The calculations are based on the NLO19(500) YN potential.
 The SRG-induced YNN interaction is also included in the calculations 
 for $^4_{\Lambda}\mathrm{He}$-$^4_{\Lambda}\mathrm{H}$ whereas 
 the $A=7$ and $8$ wave functions were computed at the magic SRG-flow parameter of $\lambda_{magic}=0.823 $~fm\textsuperscript{-1}. The singlet and triplet $\Lambda p \, (\Lambda n)$ scattering lengths predicted by the NLO19(500) are $a_{^1{\rm S}_0} = -2.649 (-3.202)$ fm  and $a_{^3{\rm S}_1 } = -1.580 (-1.467)$ fm.  }
\renewcommand{\arraystretch}{1.0}
\end{table}

In this appendix, we provide a brief summary of the contributions from the 
$^1S_0$ and $^3S_1$ $\Lambda \mathrm{N}$ partial waves to the expectation value of 
the corresponding YN potentials for the considered $A=4\,-\,8$ 
$\Lambda$-hypernuclei, see \ref{tab:Prob}. The weights of the respective $\La p$ and $\La n$
components are listed, too, which differ, of course, for the mirror hypernuclei
in question. Those weights, in combination with the different strengths of the 
$\Lambda n$ and $\Lambda p$ interactions in the singlet and triplet states
as manifested by the respective scattering lengths, see Table 2 in Ref.~\cite{Haidenbauer:2021wld}, determine the value 
for $\langle V_{\mathrm{YN}} \rangle$ and, in turn, also the values for 
$\Delta V_{\mathrm{YN(^1S_0)}}$ and $\Delta V_{\mathrm{YN(^3S_1)}}$ that are listed in
Tables~\ref{tab:csbA4} and \ref{tab:A7_csb}. 
Clearly, the signs of the two contributions $\Delta V_{\mathrm{YN(^1S_0)}}$ and 
$\Delta V_{\mathrm{YN(^3S_1)}}$ can be the same or the opposite, depending on the
concrete interplay realized in a specific mirror hypernucleus.

 \bibliographystyle{unsrturl}

\bibliography{bib/hyp-exp.bib,bib/ncsm.bib,bib/hyp-theory.bib,bib/nn-interactions.bib,bib/yn-interactions.bib,bib/faddeev-yakubovsky.bib,bib/srg.bib}

\begin{thebibliography}{10}

\bibitem{Dalitz:1964fu}
R.~H. Dalitz and F.~von Hippel.
\newblock {Electromagnetic $\Lambda$-$\Sigma_0$ mixing and charge symmetry for
  the $\Lambda$-Hyperon}.
\newblock {\em Phys. Lett.}, 10:153--157, 1964.
\newblock \href
  {http://dx.doi.org/10.1016/0031-9163(64)90617-1}
  {\path{doi:10.1016/0031-9163(64)90617-1}}.

\bibitem{Raymund:1964an}
M.~Raymund.
\newblock {The binding energy difference between the hypernuclides
  $^4_\Lambda$He and $^4_\Lambda$H}.
\newblock {\em Il Nuovo Cimento}, 32(3):555--587, 1964.
\newblock  \href
  {http://dx.doi.org/10.1007/BF02735882} {\path{doi:10.1007/BF02735882}}.

\bibitem{Juric:1973zq}
M.~Juri{\v{c}} et~al.
\newblock {A new determination of the binding-energy values of the light
  hypernuclei ($A \le 15$)}.
\newblock {\em Nucl. Phys.}, B52:1--30, 1973.
\newblock URL: \url{http://inspirehep.net/record/84234?ln=en}.

\bibitem{Botta:2016kqd}
E.~Botta, T.~Bressani, and A.~Feliciello.
\newblock {On the binding energy and the charge symmetry breaking in A
  \ensuremath{\leq} 16 $\Lambda$-hypernuclei}.
\newblock {\em Nucl. Phys. A}, 960:165--179, 2017.
\newblock   \href {http://dx.doi.org/10.1016/j.nuclphysa.2017.02.005}
  {\path{doi:10.1016/j.nuclphysa.2017.02.005}}.

\bibitem{Achenbach:2016ci}
Patrick Achenbach.
\newblock {Charge Symmetry Breaking in Light Hypernuclei}.
\newblock {\em Few-Body Syst.}, 58(1):17, December 2016.
\newblock   {http://dx.doi.org/10.1007/s00601-016-1178-x}
  {\path{doi:10.1007/s00601-016-1178-x}}.

\bibitem{Botta:2019has}
E.~Botta.
\newblock {Charge symmetry breaking in $s$- and $p$-shell
  $\Lambda$-hypernuclei: An updated review}.
\newblock {\em AIP Conf. Proc.}, 2130(1):030003, 2019.
\newblock \href {http://dx.doi.org/10.1063/1.5118393}
  {\path{doi:10.1063/1.5118393}}.

\bibitem{Davis:2005npa}
D.H. Davis.
\newblock 50 years of hypernuclear physics.
\newblock {\em Nuclear Physics A}, 754:3 -- 13, 2005.

\bibitem{HypernuclearDataBase}
P.~Eckert, P.~Achenbach, et~al.
\newblock Chart of hypernuclides --- {H}ypernuclear structure and decay data,
  2021.
\newblock
  \href{https://hypernuclei.kph.uni-mainz.de}{https://hypernuclei.kph.uni-mainz.de}.

\bibitem{Rijken:1999fc}
T~A Rijken, V~G~J Stoks, and Y~Yamamoto.
\newblock {Soft-core hyperon nucleon potentials}.
\newblock {\em Phys. Rev. C}, 59:21--40, 1999.
\newblock URL: \url{http://dx.doi.org/10.1103/PhysRevC.59.21}, \href
  {http://dx.doi.org/10.1103/PhysRevC.59.21}
  {\path{doi:10.1103/PhysRevC.59.21}}.

\bibitem{Nogga:2001ef}
A.~Nogga, H.~Kamada, and W.~Gl{\"o}ckle.
\newblock {The Hypernuclei $^4_\Lambda$He and $^4_{\Lambda}$He: Challenges for
  modern hyperon nucleon forces}.
\newblock {\em Phys. Rev. Lett.}, 88:172501, 2002.
\newblock  \href
  {http://dx.doi.org/10.1103/PhysRevLett.88.172501}
  {\path{doi:10.1103/PhysRevLett.88.172501}}.

\bibitem{Bedjidian:1979ih}
M.~Bedjidian et~al.
\newblock {Further Investigation of the Gamma-Transitions in $^4_{\Lambda}$H
  and $^4_{\Lambda}$He Hypernuclei}.
\newblock {\em Phys. Lett. B}, 83:252--256, 1979.
\newblock
  \href {http://dx.doi.org/10.1016/0370-2693(79)90697-X}
  {\path{doi:10.1016/0370-2693(79)90697-X}}.

\bibitem{Brandenburg:1978ldq}
R.~A. Brandenburg, S.~A. Coon, and P.~U. Sauer.
\newblock {Nuclear charge asymmetry in the a = 3 nuclei}.
\newblock {\em Nucl. Phys. A}, 294:305--320, 1978.
\newblock \href {http://dx.doi.org/10.1016/0375-9474(78)90220-8}
  {\path{doi:10.1016/0375-9474(78)90220-8}}.

\bibitem{Hiyama:2009ki}
E.~Hiyama, Y.~Yamamoto, T.~Motoba, and M.~Kamimura.
\newblock {Structure of A=7 iso-triplet $\Lambda$ hypernuclei studied with the
  four-body model}.
\newblock {\em Phys. Rev. C}, 80:054321, 2009.
\newblock  \href {http://dx.doi.org/10.1103/PhysRevC.80.054321}
  {\path{doi:10.1103/PhysRevC.80.054321}}.

\bibitem{Hiyama:2012sq}
E.~Hiyama and Y.~Yamamoto.
\newblock {Structure of $^{10}_{\Lambda}$Be and $^{10}_{\Lambda}$B hypernuclei
  studied with the four-body cluster model}.
\newblock {\em Prog. Theor. Phys.}, 128:105--124, 2012.
\newblock   \href {http://dx.doi.org/10.1143/PTP.128.105}
  {\path{doi:10.1143/PTP.128.105}}.

\bibitem{Hiyama:2013owa}
E.~Hiyama.
\newblock {Four-body structure of light $\Lambda$ hypernuclei}.
\newblock {\em Nucl. Phys. A}, 914:130--139, 2013.
\newblock \href {http://dx.doi.org/10.1016/j.nuclphysa.2013.05.011}
  {\path{doi:10.1016/j.nuclphysa.2013.05.011}}.

\bibitem{Gogami:2016jb}
T.~Gogami et~al.
\newblock {Spectroscopy of the neutron-rich hypernucleus $^{7}_{\Lambda}$He
  from electron scattering}.
\newblock {\em Phys. Rev.}, C94:021302, 2016.
\newblock URL: \url{http://link.aps.org/doi/10.1103/PhysRevC.93.034314}.

\bibitem{Yamamoto:2015avw}
T.O. Yamamoto et~al.
\newblock {Observation of Spin-Dependent Charge Symmetry Breaking in $\Lambda
  N$ Interaction: Gamma-Ray Spectroscopy of $^4_{\Lambda }$He}.
\newblock {\em Phys. Rev. Lett.}, 115(22):222501, 2015.
\newblock  \href {http://dx.doi.org/10.1103/PhysRevLett.115.222501}
  {\path{doi:10.1103/PhysRevLett.115.222501}}.

\bibitem{Esser:2015fm}
A.~Esser et~al.
\newblock {Observation of $^4_{\Lambda}$H Hyperhydrogen by Decay-Pion
  Spectroscopy in Electron Scattering}.
\newblock {\em Phys. Rev. Lett.}, 114(23):232501, 2015.
\newblock   \href {http://dx.doi.org/10.1103/PhysRevLett.114.232501}
  {\path{doi:10.1103/PhysRevLett.114.232501}}.

\bibitem{Schulz:2016dt}
F.~Schulz et~al.
\newblock {Ground-state binding energy of $^4_{\Lambda}$H from high-resolution
  decay-pion spectroscopy}.
\newblock {\em Nucl. Phys. A}, 954:149--160, 2016.
\newblock  \href
  {http://dx.doi.org/10.1016/j.nuclphysa.2016.03.015}
  {\path{doi:10.1016/j.nuclphysa.2016.03.015}}.

\bibitem{Gazda:2016ir}
D.~Gazda and A.~Gal.
\newblock {Charge symmetry breaking in the A=4 hypernuclei}.
\newblock {\em Nucl. Phys. A}, 954:161--175, 2016.
\newblock URL:
  \url{http://linkinghub.elsevier.com/retrieve/pii/S0375947416301294}, \href
  {http://dx.doi.org/10.1016/j.nuclphysa.2016.05.015}
  {\path{doi:10.1016/j.nuclphysa.2016.05.015}}.
\bibitem{Haidenbauer:2013oca}
J.~Haidenbauer, S.~Petschauer, N.~Kaiser, U.-G. Mei{\ss}ner, A.~Nogga, and
  W.~Weise.
\newblock {Hyperon-nucleon interaction at next-to-leading order in chiral
  effective field theory}.
\newblock {\em Nucl. Phys. A}, 915:24--58, 2013.
\newblock  \href {http://dx.doi.org/10.1016/j.nuclphysa.2013.06.008}
  {\path{doi:10.1016/j.nuclphysa.2013.06.008}}.

\bibitem{Haidenbauer:2019boi}
J.~Haidenbauer, U.-G. Mei{\ss}ner, and A.~Nogga.
\newblock {Hyperon--nucleon interaction within chiral effective field theory
  revisited}.
\newblock {\em Eur. Phys. J. A}, 56(3):91, 2020.
\newblock  \href {http://dx.doi.org/10.1140/epja/s10050-020-00100-4}
  {\path{doi:10.1140/epja/s10050-020-00100-4}}.



\bibitem{Haidenbauer:2021wld}
J.~Haidenbauer, U.-G. Mei\ss{}ner, and A.~Nogga.
\newblock {Constraints on the $\varLambda $-Neutron Interaction from Charge
  Symmetry Breaking in the $\mathbf {^4_\Lambda \mathrm{He}}$ - $\mathbf
  {^4_\Lambda \mathrm{H}}$ Hypernuclei}.
\newblock {\em Few Body Syst.}, 62(4):105, 2021.
\newblock   \href {http://dx.doi.org/10.1007/s00601-021-01684-3}
  {\path{doi:10.1007/s00601-021-01684-3}}.

\bibitem{Wirth:2014ko}
R.~Wirth, D.~Gazda, P.~Navr{\'a}til, A.~Calci, J.~Langhammer, and R.~Roth.
\newblock {Ab Initio Description of p-Shell Hypernuclei}.
\newblock {\em Phys. Rev. Lett.}, 113:192502, 2014.
\newblock   \href {http://dx.doi.org/10.1103/PhysRevLett.113.192502}
  {\path{doi:10.1103/PhysRevLett.113.192502}}.

\bibitem{Wirth:2017bpw}
R.~Wirth, D.~Gazda, P.~Navrátil, and R.~Roth.
\newblock {Hypernuclear No-Core Shell Model}.
\newblock {\em Phys. Rev. C}, 97(6):064315, 2018.
\newblock  \href {http://dx.doi.org/10.1103/PhysRevC.97.064315}
  {\path{doi:10.1103/PhysRevC.97.064315}}.

\bibitem{Wirth:2019cpp}
R.~Wirth and R.~Roth.
\newblock {Similarity renormalization group evolution of hypernuclear
  Hamiltonians}.
\newblock {\em Phys. Rev. C}, 100(4):044313, 2019.
\newblock   \href {http://dx.doi.org/10.1103/PhysRevC.100.044313}
  {\path{doi:10.1103/PhysRevC.100.044313}}.

\bibitem{Le:2020zdu}
H.~Le, J.~Haidenbauer, U.-G. Mei\ss{}ner, and A.~Nogga.
\newblock {Jacobi no-core shell model for $p$-shell hypernuclei}.
\newblock {\em Eur. Phys. J. A}, 56(12):301, 2020.
\newblock  \href {http://dx.doi.org/10.1140/epja/s10050-020-00314-6}
  {\path{doi:10.1140/epja/s10050-020-00314-6}}.

\bibitem{STAR:2022zrf}
Mohamed Abdallah et~al.
\newblock {Measurement of $\rm ^4_{\Lambda}H$ and $\rm ^4_{\Lambda}He$ binding
  energy in Au+Au collisions at $\sqrt{s_\mathrm{NN}}$ = 3 GeV}.
\newblock {\em Phys. Lett. B}, 834:137449, 2022.
\newblock  \href {http://dx.doi.org/10.1016/j.physletb.2022.137449}
  {\path{doi:10.1016/j.physletb.2022.137449}}.

\bibitem{Reinert:2018ip}
P~Reinert, Hermann Krebs, and Evgeny Epelbaum.
\newblock {Semilocal momentum-space regularized chiral two-nucleon potentials
  up to fifth order}.
\newblock {\em Eur. Phys. J. A}, 54(5):86, May 2018.
\newblock \href
  {http://dx.doi.org/10.1140/epja/i2018-12516-4}
  {\path{doi:10.1140/epja/i2018-12516-4}}.

\bibitem{Gazda:2022fte}
D.~Gazda, T.~Yadanar Htun, and C.~Forss\'en.
\newblock {Nuclear physics uncertainties in light hypernuclei}.
\newblock {\em Phys. Rev. C}, 106(5):054001, 2022.
\newblock \href {http://arxiv.org/abs/2208.02176} {\path{arXiv:2208.02176}},
  \href {http://dx.doi.org/10.1103/PhysRevC.106.054001}
  {\path{doi:10.1103/PhysRevC.106.054001}}.

\bibitem{Wirth:2018ho}
R.~Wirth and R.~Roth.
\newblock {Light neutron-rich hypernuclei from the importance-truncated no-core
  shell model}.
\newblock {\em Phys. Lett. B}, 779:336--341, April 2018.
\newblock URL:
  \url{http://linkinghub.elsevier.com/retrieve/pii/S0370269318301230}, \href
  {http://dx.doi.org/10.1016/j.physletb.2018.02.021}
  {\path{doi:10.1016/j.physletb.2018.02.021}}.
\bibitem{LENPIC:2022cyu}
P.~Maris et~al.
\newblock {Nuclear properties with semilocal momentum-space regularized chiral
  interactions beyond N2LO}.
\newblock 6 2022.
\newblock \href {http://arxiv.org/abs/2206.13303} {\path{arXiv:2206.13303}}.

\bibitem{Nuclearandhypernuc:2001wd}
A.~Nogga.
\newblock {\em {Nuclear and hypernuclear three- and four-body bound states}}.
\newblock PhD thesis, Bochum University, 2001.
\newblock URL:
  \url{https://hss-opus.ub.ruhr-uni-bochum.de/opus4/frontdoor/deliver/index/docId/3778/file/diss.pdf
}.

\bibitem{Le:2019gjp}
H.~Le, J.~Haidenbauer, U.-G. Mei{\ss}ner, and A.~Nogga.
\newblock {Implications of an increased $\Lambda$-separation energy of the
  hypertriton}.
\newblock {\em Phys. Lett. B}, 801:135189, 2020.
\newblock  \href {http://dx.doi.org/10.1016/j.physletb.2019.135189}
  {\path{doi:10.1016/j.physletb.2019.135189}}.



\bibitem{Le:2022jvc}
H.~Le.
\newblock {Single- \& double-strangeness hypernuclei up to $ A=8 $ within
  chiral effective field theory}.
\newblock In {\em {14th International Conference on Hypernuclear and Strange Particle Physics, EPJ Web of Conferences 271, 01004 (2022)}}.
\newblock \href {http://arxiv.org/abs/2210.02860} {\path{arXiv:2210.02860}}.

\bibitem{Petschauer:2016ho}
S.~Petschauer, N.~Kaiser, J.~Haidenbauer, U.-G. Mei{\ss}ner, and W.~Weise.
\newblock {Leading three-baryon forces from SU(3) chiral effective field
  theory}.
\newblock {\em Phys. Rev. C}, 93(1):3, January 2016.
\newblock  \href
  {http://dx.doi.org/10.1103/PhysRevC.93.014001}
  {\path{doi:10.1103/PhysRevC.93.014001}}.

\bibitem{Gal:2015bfa}
A.~Gal.
\newblock {Charge symmetry breaking in $\Lambda$ hypernuclei revisited}.
\newblock {\em Phys. Lett.}, B744:352--357, 2015.

\bibitem{Gal:HYP2015}
A.~Gal.
\newblock Charge symmetry breaking in {$\Lambda$} hypernuclei: Updated hyp 2015
  progress report.
\newblock {\em JPS Conf. Proc.}, 17:011006, 2017.
\newblock \href
  {http://dx.doi.org/10.7566/JPSCP.17.011006}
  {\path{doi:10.7566/JPSCP.17.011006}}.

\bibitem{Hiyama:2002yj}
E.~Hiyama, M.~Kamimura, T.~Motoba, T.~Yamada, and Y.~Yamamoto.
\newblock {Four-body cluster structure of A = 7 -10 double Lambda hypernuclei}.
\newblock {\em Phys. Rev. C}, 66:024007, 2002.
\newblock \href
  {http://dx.doi.org/10.1103/PhysRevC.66.024007}
  {\path{doi:10.1103/PhysRevC.66.024007}}.

\end{thebibliography}

\end{document}